\documentclass[preprint,prd,tightenlines,nofootinbib]{revtex4}
\usepackage{graphicx}
\def\DESepsf(#1 width #2){\includegraphics[width = #2]{#1}}
\begin{document}

\title{Partons and Jets at the LHC}
\author{Davison E.~Soper}
\affiliation{University of Oregon}

\begin{abstract}
I review some issues related to short distance QCD and
its relation to the experimental program of the Large Hadron Collider
(LHC) now under construction in Geneva.
\end{abstract}

\maketitle

\section{Introduction}

The tools of Quantum Chromodynamics (QCD) will be important in planning
and analyzing experiments at the Large Hadron Collider (LHC). I review
some of the issues here. In particular, I discuss some methods for
finding new physics at the LHC in ``generic'' searches. Then I discuss
the sources of theory error. Finally, I review the state of the art in
higher order calculations that are important for confronting the
experimental results with theory.  

\section{How to find new physics at the LHC.}

\begin{figure}[htb]
\centerline{\DESepsf(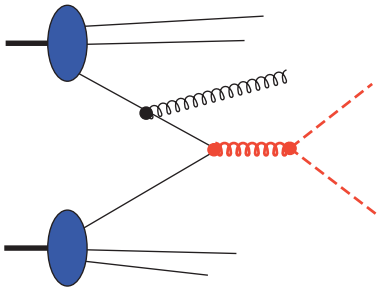 width 5 cm)}
\caption{A diagram for  
$p+ p \to {\rm squark} + {\rm antisquark} + X$. The squark and antisquark
are created via their coupling to a gluon.}
\label{fig:squarks}
\end{figure}

The most obvious way to look for new physics at the LHC is to look
directly. For instance, one can look for supersymmetry by looking for the
process $p+ p \to {\rm squark} + {\rm antisquark} + X$, as depicted in
Fig.~\ref{fig:squarks}.

A calculation of the cross section for this process requires QCD tools
\cite{QCDhandbook}. One starts with the factorization formula
\begin{equation}
{{d \sigma}} \approx\! 
\sum_{a,b} \int_{0}^1\! d \xi_A \int_{0}^1\! d \xi_B\
{{f_{a/A}(\xi_A,\mu)}}\ {{f_{b/B}(\xi_B,\mu)}}\
{{{{d \hat\sigma^{ab}(\mu) }}}}.
\label{factorization}
\end{equation}
One needs the parton distribution functions
${{f_{a/A}(\xi,\mu)}}$ to tell the probability to find the required
initial partons in the protons. We also need the hard scattering cross
sections ${{{{d \hat\sigma^{ab}(\mu) }}}}$ for these partons to produce
the squark and antisquark. This diagram illustrates a piece of the
next-to-leading order (NLO) calculation of ${{{{d \hat\sigma^{ab} }}}}$.
The emitted gluon could be included in ${{{{d \hat\sigma^{ab} }}}}$ or it
could be included in the parton distribution functions. The calculation
of ${{{{d \hat\sigma^{ab} }}}}$ contains a subtraction term so as not to
count the gluon emission twice. Calculations are available at the NLO
level for a wide variety of new physics signals of interest.

\begin{figure}[htb]
\centerline{\DESepsf(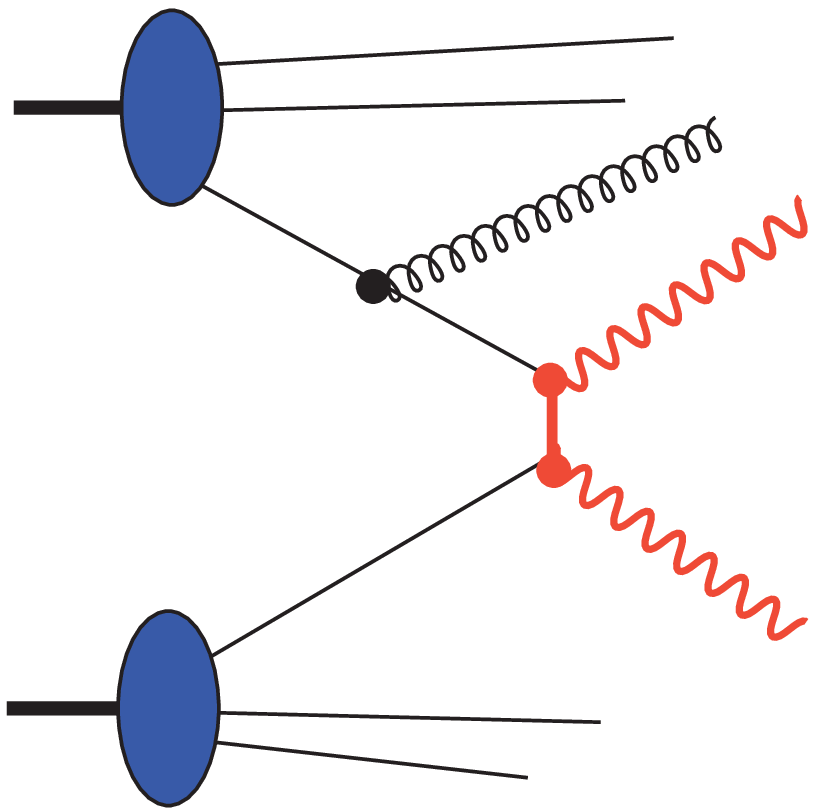 width 5 cm)}
\caption{A diagram for  $p + p \to W^+ + W^- + X$.}
\label{fig:makeW}
\end{figure}

One can also use indirect methods to look for new physics at the LHC. For
example, consider the Standard Model process $p + p \to W^+ + W^- + X$,
as depicted in Fig.~\ref{fig:makeW}. We still use
Eq.~(\ref{factorization}) and perform a NLO calculation of the cross
section for this process. If the experimental result does not agree with
the theory, there must be new physics (if the discrepancy is outside
the errors). Evidently, for this purpose one needs accurate calculations.
Next-to-next-to-leading order would be nice (but is not available). Also,
one needs a serious estimate of the errors.

\section{Jet cross sections and new physics signatures}

Another indirect method to look for new physics at the LHC is to look
for the cross section to make two jets, $p + p \to {\rm jet} + {\rm jet} +
X$ (or, almost equivalently, one jet plus anything). Here a jet is a
spray of particles. The exact definition is important, but we leave it
aside for the moment. If we measure the two-jet cross section, then it is
useful to measure the invariant mass of the two-jet system $M_{JJ}$. If
we simply measure one jet, then a good variable to use is the so-called
transverse energy $E_T$ of the measured jet. This is the sum of the
absolute values of the transverse momenta of the particles in the jet.

\begin{figure}[htb]
\centerline{\DESepsf(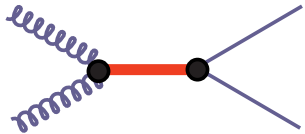 width 4 cm)}
\caption{Production of a new particle that decays to two jets.}
\label{fig:newtwo}
\end{figure}

To see why jet cross sections are a useful tool for looking for new
physics, consider the possibility that there is some sort of new
interaction at scale $\Lambda$ and that $\Lambda$ is less than the
maximum value of the parton-parton c.m.\ energy available in the
experiment ({\it i.e.} a few TeV for the LHC). One possibility is that
there is a new particle with mass $M$ that decays to two jets, as
depicted in  Fig.~\ref{fig:newtwo}. To see this in the one jet inclusive
cross section, one would look for a threshold at $E_T = M/2$. Above this
threshold, the cross section would be bigger than predicted in the
Standard Model. To see this in the two-jet inclusive cross section, one
would look for a resonance peak at $M_{JJ} = M$. That would be a
spectacular signal.

\begin{figure}[htb]
\centerline{\DESepsf(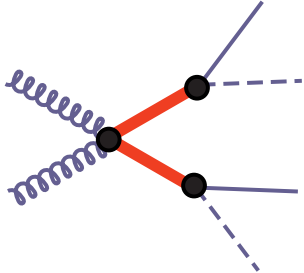 width 4 cm)}
\caption{Production of new particles that decay to two jets and two
leptons.}
\label{fig:newfour}
\end{figure}

There are other possibilities. For instance, suppose that there is a new
particle of mass $M$ that is produced in pairs. If each particle decays
to two jets, one could see a threshold effect in the four-jet cross
section (which, unfortunately, is known only to leading order). On the
other hand, suppose that each particle decays to a lepton and a jet, as
depicted in Fig.~\ref{fig:newfour}. In principle, this process would
contribute to the one jet and two-jet inclusive cross sections. However,
the signal would likely be much smaller than the background. That is
because jet cross sections fall steeply with increasing $E_T$ because it
is rare to find two incoming partons with large c.m.\ energy. When we
look for just the jets and not the leptons, we miss seeing much of the
c.m.\ energy in the event, so we bin the event together with events that
have lower c.m.\ energy and thus higher cross section. The result is that
to see the type of process depicted in Fig.~\ref{fig:newfour}, one needs
to look for the leptons in addition to the jets. 

Now suppose that there is some sort of new interaction at scale $\Lambda$
and that $\Lambda$ is {\it greater} than the maximum value of
parton-parton c.m.\ energy available in the experiment. Now searching for
the new interaction seems hopeless. However, it is not hopeless, because
the new interaction will lead to new terms in the effective lagrangian
such as, for example,
\begin{equation}
\Delta {\cal L} = {g'\over \Lambda^2}\, (\bar\psi \psi)^2.
\label{effL}
\end{equation}
Here $g'$ is a dimensionless coupling that arises from the new
interactions, while $\Lambda$ is a scale parameter with dimension $m^1$.
The factor ${1/ \Lambda^2}$ is present for dimensional reasons: $\Delta
{\cal L}$ has dimension $m^4$ and $(\bar\psi \psi)^2$ has dimension $m^6$.
This term in the effective lagrangian modifies the jet cross section:
\begin{equation}
{d \sigma_{\rm Jet} \over dE_T}
\approx
\left(d \sigma_{\rm Jet} \over dE_T\right)_{\!\!0}
\times\left[
1 + ({\rm const.})\, {g'\over \alpha_s}\,{E_T^2 \over \Lambda^2}
\right].
\end{equation}
Here the cross section with subscript 0 is the Standard Model cross
section. The extra contribution depicted arises from the interference
between the Standard Model amplitude and the $\Delta {\cal L}$ amplitude.
There is a factor ${g' / \Lambda^2}$ because that factor is present in
$\Delta {\cal L}$. There is a factor $1/\alpha_s$ because we are
replacing a Standard Model amplitude that has a factor $\alpha_s$.
Finally, there is a factor $E_T^2$ to make the dimensions right.

Now the strategy is evident. For small $E_T$, the addition to the cross
section is negligible. However, the addition grows with $E_T$. Thus we
should look for a discrepancy between theory and experiment that grows
with $E_T$.

Here is one more possibility. What if space has more than three
dimensions, with the extra dimensions rolled into a little ball of size
$R$? Then a quark or gluon is pointlike when viewed by a probe with
wavelength $\lambda \gg R$, but not when viewed by a probe with 
wavelength $\lambda \lesssim R$. Then the one jet inclusive cross section
should be suppressed by a form factor something like
\cite{Oda}
\begin{equation}
{d \sigma_{\rm Jet} \over dE_T}
\approx
\left(d \sigma_{\rm Jet} \over dE_T\right)_{\!\!0}
\times \exp(- R E_T).
\end{equation}
That would be a spectacular signal.

\begin{figure}[htb]
\centerline{\DESepsf(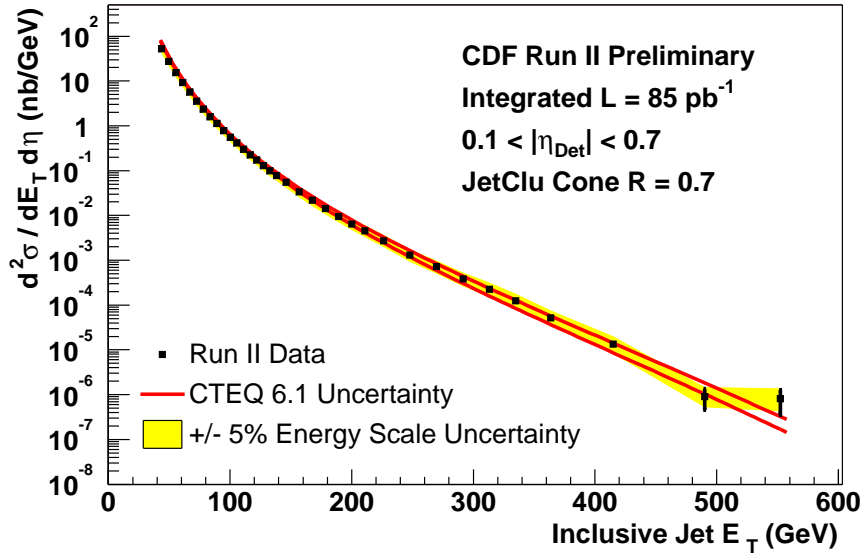 width 12 cm)}
\caption{Preliminary Run II inclusive jet cross section from
CDF\protect\cite{CDFtalk}.}
\label{fig:CDFjets}
\end{figure}

I display in Fig.~\ref{fig:CDFjets} the latest results \cite{CDFtalk} from
the CDF experiment at Fermilab on the inclusive jet cross section. These
results are from Run II of the Tevatron, which is currently underway.
These results extend further in $E_T$ than the results obtained in the
previous run. The experimental measurement is compared to the
prediction from next-to-leading order QCD. Theory and experiment agree
within the uncertainties, indicating that ``new physics'' effects remain
outside the range of the experiment.

\begin{figure}[htb]
\centerline{\DESepsf(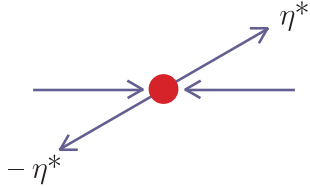 width 4 cm)}
\caption{Two jet production as viewed in the jet-jet c.m.\ frame. The
rapidity $\eta^*$ is related to the scattering angle by $\eta^*
= - \ln\tan(\Theta^*/2)$.}
\label{fig:twojetangle}
\end{figure}

One can also measure the two-jet inclusive cross section. For each
event, find the two jets with the largest $E_T$ and use these to
define the cross section
\begin{equation}
{d\,\sigma \over d\,M_{JJ}\ d\,\eta _{JJ}\ d\,\eta^*} \ .
\end{equation}
where $\eta_{JJ} = (\eta_1 + \eta_2)/2$ is the rapidity of the jet-jet
c.m. system and $ \eta^* = (\eta_1 - \eta_2)/2$ is the rapidity of the
first jet as viewed in the jet-jet c.m. system. This is illustrated in
Fig.~\ref{fig:twojetangle}. If we integrate over the rapidities to form
$d\,\sigma / d\,M_{JJ}$, we get a cross section that contains essentially
the same information as the one jet inclusive cross section $d\,\sigma /
d\,E_T$, except that a resonance that decays to two jets would appear as
a bump in the plot of $d\,\sigma / d\,M_{JJ}$ versus $M_{JJ}$. If,
however, we look at the angular distribution $[d\sigma/d\,M_{JJ}\
d\,\eta^*]/[d\sigma/d\,M_{JJ}]$, we get new information. That is because
vector boson exchange, which is characteristic of QCD, gives the
characteristic behavior
\begin{equation}
{d\,\sigma \over \ d\,\eta^*} \propto \exp (2 \eta^*) 
\hskip 1.0 cm
\eta^* \gg 1 \ .
\end{equation}
In contrast, an s-wave distribution gives few events with $ \eta^* >1$.
More generally, a distribution with any small value of orbital angular
momentum in the $s$-channel, which would be characteristic of
a new physics signal,  gives few events with $ \eta^* >1$. A convenient
angle variable is $\chi =  \exp(2 \eta^*)$,
which gives
\begin{equation}
{d\,\sigma \over \ d\,\chi} 
= {1 \over 2\,\exp(2 \eta^*)}{d\,\sigma \over \ d\,\eta^*}.
\end{equation}
The QCD cross section is quite flat for $\chi \gg 1$. In contrast, a
new physics signal should fall off beyond $\chi \approx 3$. This
method has been used for the analysis of Fermilab data. As with the
one jet inclusive cross section, evidence for new physics has not been
seen \cite{CDFdijet}.

\begin{figure}[htb]
\centerline{\DESepsf(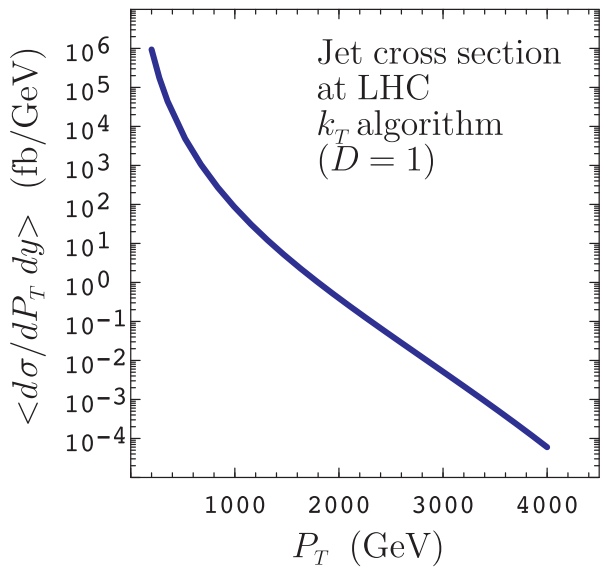 width 8 cm)}
\caption{QCD prediction at next-to-leading order for the one jet
inclusive cross section at the LHC. I plot the cross section
$d\sigma/dP_T\,dy$ averaged over $-1<y<1$, versus $P_T$. The jet
definition used is the successive combination jet definition, $k_T$
style, with joining parameter $D=1$\protect\cite{ktjetdef}. The
renormalization and factorization scale are $\mu_{\rm UV} = \mu_{\rm
coll} = P_T/2$. I use CTEQ5M parton distributions \protect\cite{CTEQ5}.
The code used is from
\protect\cite{EKSjets}.}
\label{fig:LHCjets}
\end{figure}

Thus we look forward to the results at much higher $E_T$ or $M_{JJ}$ that
will come from the LHC. In Fig.~\ref{fig:LHCjets}, I display the QCD
prediction.

\section{Theory errors}

What are the errors in the theoretical prediction? Consider the
example of the one jet inclusive cross section. The cross section
starts with contributions of order $\alpha_s^B$ with $B = 1$. We know
the cross section to next-to-leading order, which means that terms of
order $\alpha_s^{B+1}$ are also included. However, terms of order
$\alpha_s^{B+2}$ and higher are not included. To get an estimate of
the theory error arising from not including these terms, one can argue
that the omitted terms are not likely to be smaller than certain order 
$\alpha_s^{B+2}$ terms that we know about. These are terms
\begin{eqnarray}
&&\alpha_s^{B+2} \times
[
C^{(1)}_{\rm UV}\log(2\mu_{\rm UV}/E_T)
+
C^{(1)}_{\rm coll}\log(2\mu_{\rm coll}/E_T)
\nonumber\\
&&
+
C^{(2)}_{\rm UV}\log^2(2\mu_{\rm UV}/E_T)
+
C^{(2)}_{\rm coll}\log^2(2\mu_{\rm coll}/E_T)
\nonumber\\
&&
+C^{(2)}_{\rm mixed}\log(2\mu_{\rm UV}/E_T)\
\log(2\mu_{\rm coll}/E_T)
],
\label{scales}
\end{eqnarray}
where $\mu_{\rm UV}$ is the renormalization scale
and $\mu_{\rm coll}$ is the factorization scale. Our
hypothesis is that the unknown $\alpha_s^{B+2}$ terms are not likely
to be smaller than these terms, with the logarithms replaced by
something of order 1 ($\pm\log 2$ is the conventional choice).

It is easy to estimate these terms. Consider for instance
the dependence on $\mu_{\rm UV}$. The scale $\mu_{\rm UV}$ occurs in the
argument of $\alpha_s$. We know that the derivative of the cross section
with respect to $\mu_{\rm UV}$ would be zero if the cross section were
calculated exactly. However the terms in  Eq.~(\ref{scales}) are absent.
Thus the derivative of the calculated cross section with respect to
$\mu_{\rm UV}$ is just the derivative of these terms plus higher order
terms. 

This leads to a simple and widely used method for estimating the theory
error. We first choose a nominal ``best'' value of the scales, say
$E_T/2$. Then we vary the scales by a factor of 2 about this choice.
The change in the calculated cross section under this variation
provides the error estimate.

\begin{figure}[htb]
\centerline{\DESepsf(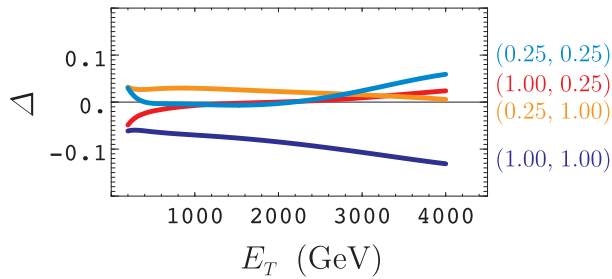 width 8 cm)}
\caption{Variation of the jet cross section with the choice of
renormalization and factorization scales. The fractional change in
the cross section is shown as a function of $E_T$ for four different
choices for $(\mu_{\rm UV},\mu_{\rm coll})$.}
\label{fig:jeterrors}
\end{figure}

Let us try this for the case of the jet cross section at the LHC.
Define
\begin{equation}
\Delta(\mu_{\rm coll},\mu_{\rm UV})
= 
{d\sigma(\mu_{\rm UV},\mu_{\rm coll})/dE_T
\over
d\sigma(E_T/2,E_T/2)/dE_T
}
-1.
\end{equation}
Then we can plot $\Delta$ versus $E_T$ for four choices of the scales
$(\mu_{\rm UV},\mu_{\rm coll})$,
\begin{equation}
(E_T/4,E_T/4),(E_T,E_T/4),(E_T/4,E_T),(E_T,E_T).
\end{equation}
The results are shown in Fig.~\ref{fig:jeterrors}. We see that the
perturbative errors could reasonably be estimated at $\pm 10\%$.

There is another class of theory errors. These are the errors inherent
in using the factorization theorem, Eq.~(\ref{factorization}), and are
of the form
\begin{equation}
{d\sigma\over dE_T} = \left({d\sigma\over dE_T}\right)_{\rm NLO}
\left\{
1 + {\Lambda_1 \over E_T} + {\Lambda_2^2 \over E_T^2} + \cdots
\right\}.
\end{equation}
A rough estimate is that the $\Lambda_i$ are of order $10\ {\rm GeV}$
for experiments at the Tevatron, with $\sqrt s \approx 2\ {\rm TeV}$,
and perhaps somewhat larger for the LHC. This estimate arises from
supposing that a jet can gain a GeV or more of $E_T$ from energy in the
event that did not come from the hard collision, or it could loose a
GeV of $E_T$ that falls outside of the jet cone. Then the rapid
decrease of the cross section with $E_T$, something like $E_T^{-8}$
gives an estimate for $\Lambda_1$ of order 10.

These power suppressed terms may be important for some of the Tevatron
experiments involving jets with $E_T < 50\ {\rm GeV}$, but they are not
likely to be important for jets with $E_T > 200\ {\rm GeV}$ at the LHC.

There is one more obvious source of error in the theoretical
prediction. That arises in the parton distributions that appear in
Eq.~(\ref{factorization}). For momentum faction $x < 0.3$, I suppose
that we know the parton distributions (which are derived from
experimental results) to $\pm 10\%$ or better. This would give a 
$\pm 20\%$ error on LHC cross sections.

\begin{figure}[htb]
\centerline{\DESepsf(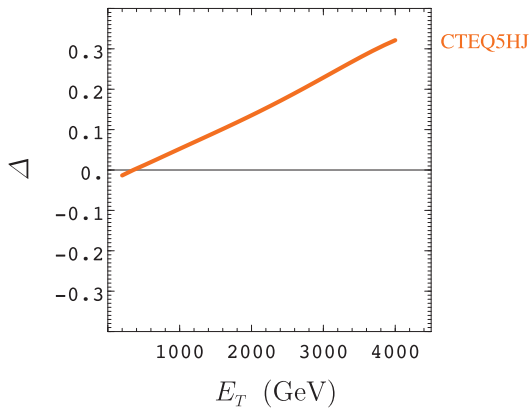 width 8 cm)}
\caption{Jet cross section at LHC calculated with CTEQ5HJ partons
compared to the cross section calculated with CTEQ5M partons
\protect\cite{CTEQ5}. The
fractional difference between the calculated cross sections is
plotted versus $E_T$.}
\label{fig:jetswith5HJ}
\end{figure}

For larger $x$, our knowledge of the gluon distribution is poor. To
take an example, I compare the jet cross section calculated with
$CTEQ5HJ$ partons with that calculated with $CTEQ5M$ partons \cite{CTEQ5}.
The primary difference between these two sets lies in the large $x$
gluons. Both provide reasonable fits to the experimental data on which
they are based. In Fig.~\ref{fig:jetswith5HJ}, I plot
\begin{equation}
\Delta
= 
{d\sigma(CTEQ5HJ)/dE_T
\over
d\sigma(CTEQ5M)/dE_T
}
-1
\end{equation}
versus $E_T$. We see that the difference reaches 30\% for 4 TeV jets.

There has been an important development in the last couple of years:
parton distribution functions now come with estimated errors
\cite{CTEQ6,MRSTerrors}. This will make the estimation of errors for the
prediction of LHC cross sections more reliable.

\section{Reducing the perturbative theory error}

It should be clear that it would be good to have more accuracy
available in the theoretical predictions. To achieve this, we should
try to do calculations at NNLO, {\it i.e.} including terms of order
$\alpha_s^B$, $\alpha_s^{B+1}$, and $\alpha_s^{B+2}$ for a process
whose theoretical expression begins at order $\alpha_s^B$.

\begin{figure}[htb]
\centerline{\DESepsf(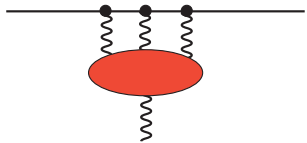 width 4 cm)}
\caption{Light-by-light scattering, which contributes to $g-2$ of the
muon.}
\label{fig:lightbylight}
\end{figure}

To illustrate the point that precision is important in science, it is
of interest to recall the case of $g - 2$ for the muon, which was a
topic of great interest last year. The ongoing experiment 
E821 at Brookhaven \cite{gminus2} (averaged with previous results) gives
\begin{equation}
(g-2)/2 = 11\,659\,203(8) \times 10^{-10}.
\end{equation}
The corresponding calculation includes QED corrections at 
N${}^4$LO, {\it i.e.} $\alpha^{B+4} =\alpha^5$. The calculation
also includes two loop graphs with $W$ and $Z$ bosons. There are also QCD
contributions, which cannot be purely perturbative because the momentum
scale is too low. One of the QCD contributions, light-by-light
scattering, is illustrated in Fig.~\ref{fig:lightbylight}. This
contribution had a sign error, fixed by Knecht and Nyffeler
\cite{KnechtNyffeler}, who found that this graph contributes
$+8.3(1.2)\times10^{-10}$. The revised theoretical contribution helps to
reduce the difference between theory and experiment.

The theory result, as summarized in \cite{Nyffeler}, is
\begin{eqnarray}
(g-2)/2 &=& 11\,659\,168(9) \times 10^{-10},
\nonumber\\
(g-2)/2 &=& 11\,659\,193(7) \times 10^{-10},
\end{eqnarray}
depending on whether one uses $e^+e^-$ or $\tau$ data, respectively, to
obtain the contributions from vacuum polarization diagrams involving
hadrons.  In view of the difference between the two theory numbers, there
may well be more theoretical uncertainty than indicated by the quoted
errors. However, there is certainly a hint of a discrepancy between
theory and experiment.  This has a bearing on LHC physics because it
suggests beyond the Standard Model physics of some sort at a mass scale
of some fraction of a TeV. (Compare Eq.~(\ref{effL}).) This discussion
also illustrates that N${}^k$LO calculations matter.

\begin{figure}[htb]
\centerline{\DESepsf(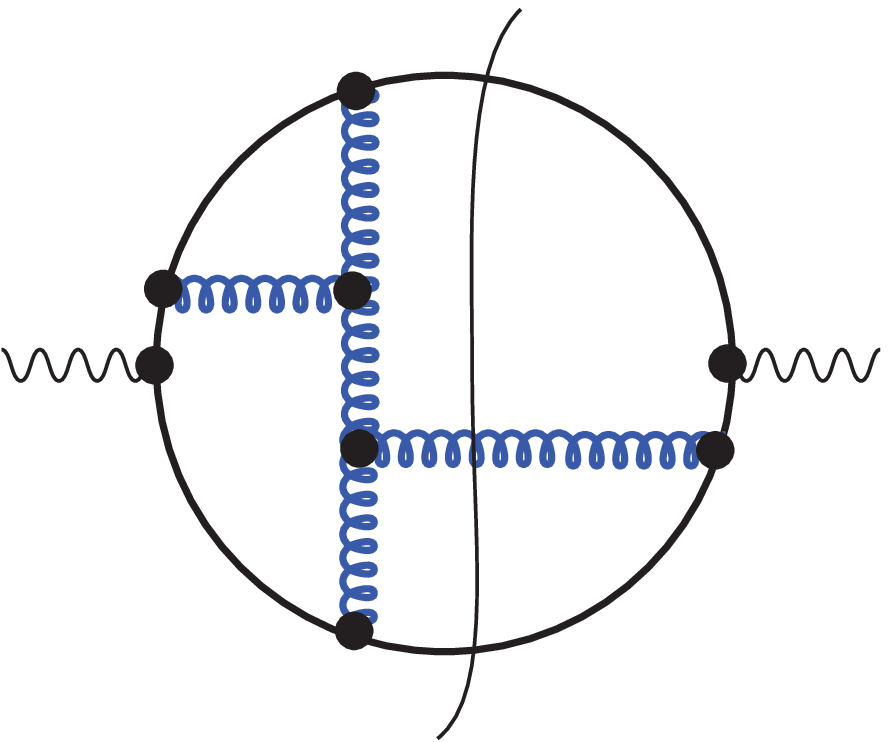 width 4 cm)
            \DESepsf(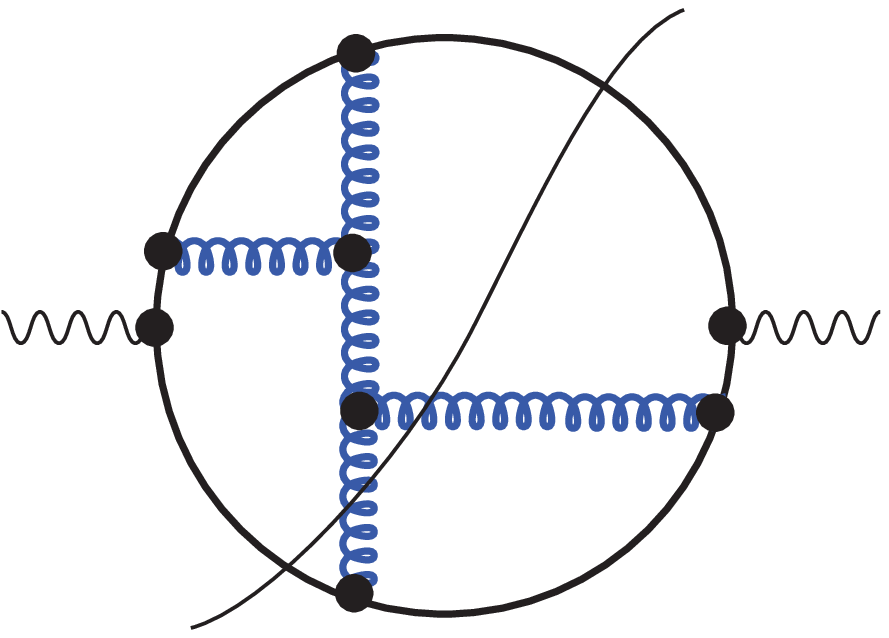 width 4 cm)}
\caption{Two cut graphs for $e^+e^- \to 3\ {\rm
jets}$ at NNLO.}
\label{fig:nnlojets}
\end{figure}

There are some calculations in QCD at NNLO. These important
calculations are successful because they use special tricks based on
calculating a simple measurable quantity. An example is the calculation
\cite{Surguladze} of the total cross section for $e^+e^-$ annihilation
to hadrons at order $\alpha_s^3$. For more complicated quantities, such
as the rapidity distribution of muon pairs produced in hadron
collisions\cite{DYatNNLO}, the calculations are harder. Yet more difficult
is to calculate for {\it generic} infrared safe observables. The first
results for generic observables will probably come for $e^+e^-
\to 3\ {\rm jets}$. In Fig.~\ref{fig:nnlojets}, we see an example of two
of the graphs that must be calculated for this purpose.

In the standard analytical/numerical method for calculating such
graphs, one needs an analytic result for the two loop virtual graph
that appears in the left hand graph in Fig.~\ref{fig:nnlojets}. This
graph is infrared divergent and is regularized by doing the
calculation in  $4-2\epsilon$ dimensions. (This particular graph also
has an ultraviolet divergent one loop subgraph, which is treated with
the aid of the same dimensional regulation.) Understandably, it is not
so easy to obtain analytical results for graphs like this.

There are also cut graphs with 4 and 5 final state partons. There
are infrared divergences that arise when one integrates over the
phase space for these partons. Ultimately, these integrations are
performed by numerical integration, but first one needs a subtraction
scheme that eliminates the infrared divergences. This is also not so
simple.

There has been recent progress. The analytical results for the two
loop virtual graphs are available. The subtraction scheme is in
progress. The progress is being made because there are several very
good people working on this. For example,
C.~Anastasiou, M.~Beneke, 
Z.~Bern, K.~G.~Chetyrkin, L.~Dixon, T.~Gehrmann,  E.~W.~Glover,
S.~Laporta, S.~Moch, C.~Oleari, E.~Remiddi, V.~A.~Smirnov,
J.~B.~Tausk, P.~Uwer, O.~L.~Veretin,  and S.~Weinzierl. For a summary of
this effort, see \cite{BernICHEP2002}.

I should also point out that at NLO, it is possible to do this kind of
calculation by a completely numerical method \cite{beowulf}. This
offers evident advantages in flexibility. Perhaps a completely
numerical method could help at NNLO. This avenue is, however, not
being pursued at the moment.

\end{document}